\documentclass[natbib]{emulateapj}

\input epsf

\def\s2n{S^{\prime}/N}

\begin{document}
\title{Can We Trust the Dust? Evidence of Dust Segregation in Molecular Clouds}
\author{Paolo Padoan\altaffilmark{1}, Laurent Cambr\'{e}sy\altaffilmark{2}, 
Mika Juvela\altaffilmark{3}, Alexei Kritsuk\altaffilmark{1}, William D.
Langer\altaffilmark{4}, Michael L. Norman\altaffilmark{1}}

\altaffiltext{1}{Department of Physics, University of California, San Diego, 
CASS/UCSD 0424, 9500 Gilman Drive, La Jolla, CA 92093-0424; ppadoan@ucsd.edu}
\altaffiltext{2}{Observatoire Astronomique de Strasbourg, 67000 Strasbourg, 
France}
\altaffiltext{3}{Helsinki University Observatory, FI-00014, University of 
Helsinki, Finland}
\altaffiltext{4}{Jet Propulsion Laboratory, 4800 Oak Grove Drive, MS 183-335,
     California Institute of Technology, Pasadena, CA 91109-8099, USA}

\begin{abstract}

Maps of estimated dust column density in molecular clouds are usually assumed 
to reliably trace the total gas column density structure. In this work we 
present results showing a clear discrepancy between the dust and the gas
distribution in the Taurus molecular cloud complex. We compute the power 
spectrum of a 2MASS extinction map of the Taurus region and find it is
much shallower than the power spectrum of a $^{13}$CO map of the same 
region previously analyzed. This discrepancy may be explained
as the effect of grain growth on the grain extinction efficiency. However,
this would require a wide range of maximum grain sizes, which is ruled out based
on constraints from the extinction curve and the available grain models.
We show that major effects due to CO formation and depletion are also ruled
out. Our result may therefore suggest the existence of intrinsic spatial fluctuations
of the dust to gas ratio, with amplitude increasing toward smaller scales. 
Preliminary results of numerical simulations of trajectories of inertial 
particles in turbulent flows illustrate how the process of clustering of dust grains 
by the cloud turbulence may lead to observable effects. However, these results cannot 
be directly applied to large scale supersonic and magnetized turbulence at present.

\end{abstract}


\section{Introduction}

The spatial structure of interstellar clouds is the result of dynamical processes
responsible for the formation and evolution of the clouds and controlling the
formation of stars. Numerical simulations of star formation, coupling turbulent 
fragmentation, self-gravity and magnetic fields, should be able to reproduce
the observed cloud structure. Maps of column densities are the most straightforward
tool for the comparison of numerical and observational data, but the limitations 
of any column density tracer in star-forming clouds must be well understood  
before attempting such a comparison.

Integrated intensity maps of the $^{13}$CO J=1-0 emission line can be used 
to estimate column densities in the range of
visual extinction $2<A_V<10$~mag, if spatial variations of temperature and line
saturation are accounted for with the help of radiative transfer modeling.
Variations of the abundance of the CO molecule at $A_V<2$~mag and $A_V>10$~mag 
may be expected because of the chemistry of CO formation and depletion respectively.
Similar considerations apply to other molecular tracers, for different 
values of visual extinction. 

FIR maps of dust emission have also been used
to estimate column density, but the method requires a careful correction for
temperature variations, which can be achieved only if various FIR bands
are available and assuming some knowledge of the structure along the line of 
sight. At sub-mm wavelengths the temperature dependence is less critical,
but, as for the FIR, the thermal emission is sensitive to the optical 
properties of dust grains, which are expected to vary with density, due to grain 
growth and ice deposition. Moreover, ground based sub-mm observations do not
provide accurate maps of very extended emission, because sky fluctuations are
hard to distinguish from smooth surface brightness variations.

Considering the difficulties of the various methods of measuring column 
densities in interstellar clouds, the NIR reddening of background stars is 
usually assumed to be the most reliable probe of column density. This method is
not affected by any gradients in dust temperature and provides a direct 
measurement of the dust column density from observations of the NIR color 
excess. The NIR color excess is converted to a dust column density via the 
NIR extinction law; the dust column density is then converted to a gas 
column density by assuming a constant gas to dust ratio. 

In this work we test the hypothesis that the dust distribution faithfully
traces the total gas column density, by comparing the power spectra of 
maps of the Taurus region obtained with the J=1-0 line of $^{13}$CO and with
2MASS extinction measurements. The $^{13}$CO study was reported in Padoan et 
al. (2004). Here we present the new 2MASS results and discuss their comparison
with the CO observations. We find that the dust power spectrum is significantly 
shallower than the gas power spectrum, even after accounting for radiative 
transfer, temperature distribution, CO formation, and CO depletion effects 
on the $^{13}$CO data and after a detailed simulation of the extinction method, 
based on three dimensional density fields from the highest resolution 
simulations of supersonic turbulence to date. 

\begin{figure}[ht]
\centerline{
\epsfxsize=7.2cm \epsfbox{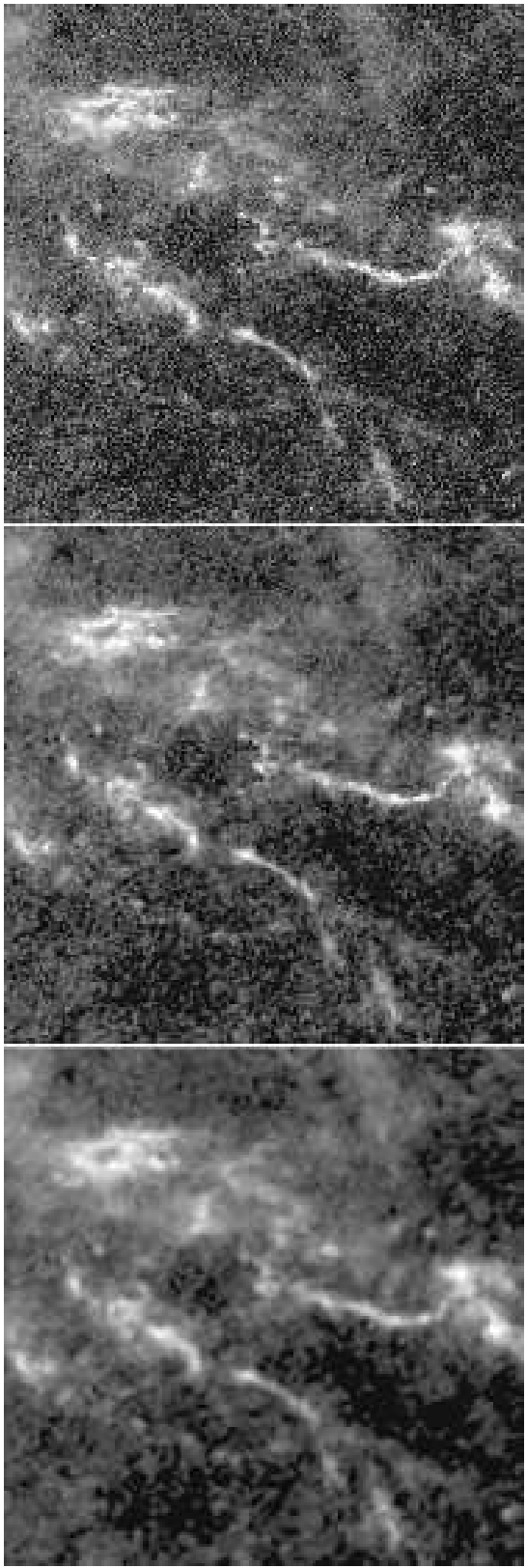}
}
\caption[]{Extinction maps of the Taurus region with 3, 10 and 30 stars per 
cell (top to bottom). The grey table is for the square root of the extinction, 
from $A_{\rm V}=0$~mag (black) to $A_{\rm V}=25.7$~mag (white). The peak 
extinction in each map is 25.7, 19.5 and 11.9~mag (top to bottom) and all 
negative values have been put equal to zero.  
}
\label{f1}
\end{figure}

Possible variations of dust extinction efficiency due to grain growth with 
increasing density may contribute to this result. However, we find that the
necessary variation in the extinction efficiency would require the maximum
grain size to span the approximate range 0.1-0.6~$\mu$m, from low to high 
extinction regions on the Taurus map. The available grain models and the
observational constraints from the extinction curve imply a smaller range
of maximum grain sizes, insufficient to explain the power spectrum discrepancy. 
The difference in the power spectra may therefore be due to a real difference 
in the structure of dust and gas column densities, with fluctuations of the 
dust to gas ratio increasing toward smaller scales. Preliminary results of the  
numerical simulation of inertial particle trajectories in turbulent flows 
support the idea that the cloud turbulence may cause a spatial clustering of 
the dust grains, resulting in the observed effect on the power spectra. 
However, the simulations are presently limited to transonic and non-magnetized 
turbulence, and so they are at best relevant only to scales $<0.1$~pc. 
Simulations of the same process with supersonic and magnetized flows are 
required for a direct interpretation of the observations.

\section{Extinction Maps and Power Spectra}

The method of computing extinction maps of interstellar clouds was 
first discussed by Lada et al. (1994) and later improved by Lombardi 
\& Alves (2001). The method is based on subdividing the observed region 
with a regular grid, and assigning to each grid cell a value of 
extinction equal to the average extinction of the stars within that cell. 
The number of stars per cell decreases with increasing average extinction in 
the cell, because only the brightest background stars can be seen through a 
large column of dust. 

In this work we adopt the method of Cambr\'{e}sy et al. (2002). This method 
takes advantage of adaptive cells with a fixed number of stars instead of cells 
of fixed size. In this way it is possible to keep the spatial resolution as high as 
allowed by the local stellar density. The spatial resolution is higher in regions of 
low extinction than in regions of large extinction, where fewer background stars are 
detected and larger cells must be used. The average spatial resolution over the whole 
map can be changed by changing the number of stars per cell. 

The color excess is computed using the relation 
$E_{H-K_{\rm s}} = (H-K_{\rm s})_{\rm obs}-(H-K_{\rm s})_{\rm int}$, 
where $(H-K_{\rm s})_{\rm obs}$ is the observed median color in a cell and 
$(H-K_{\rm s})_{\rm int}$ is the intrinsic median color, estimated from the colors 
of unreddened stars. In the method of Lada et al. (1994) the mean color is used 
instead of the median color. We use the median color because it minimizes the effect 
of foreground stars (or any object with very peculiar colors), as shown by 
Cambr\'{e}sy et al. (2002). Visual extinction values 
are obtained from the color excess using the Rieke and Lebofsky (1985) extinction 
law, which results in the relation $A_{\rm V} = 15.87 \times E_{H-K_{\rm s}}$. 

We have computed extinction maps of the Taurus region with 1, 3, 10, 30 and 100 
stars per cell. The maps with 3, 10 and 30 stars per cell (top to bottom panels) 
are shown in Figure~\ref{f1}. After removing 129 young stellar objects known from 
the literature and detected in 2MASS, the intrinsic 
color is computed as the median color of stars in regions where no $^{12}$CO 
(Dame et al. 2001) is detected, within a field larger than the actual extinction map. 
The intrinsic color is found to be $(H-K_{\rm s})_{\rm int}=0.13$~mag. 
The standard deviation of the color of these unreddened stars provides an 
estimate of the uncertainty in the extinction maps. The extinction map of the
Taurus region with 10 stars per cell has extinction values ranging from 
\begin{figure}[ht]
\centerline{
\epsfxsize=9.3cm \epsfbox{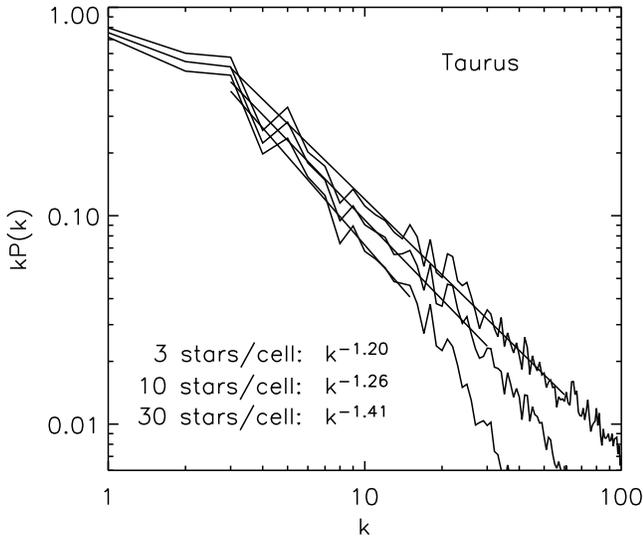}
}
\caption[]{Power spectra of the extinction maps from Figure~\ref{f1}. The maps have 
been re-sampled to the resolution of the map with 3 stars per cell. At the original
average resolution of each map, the Nyquist frequency is $k=$100, 55 and 32.
}
\label{f2}
\end{figure}
$A_{\rm V}=0.46$~mag (1--$\sigma$ detection) to $A_{\rm V}=19.5$~mag
and a mean angular resolution of $3.9\pm 0.9$~arc-minutes (the largest value
is 10.9~arc-minutes). Based on a standard gas to 
dust ratio (Bohlin et al. 1978), and a normal extinction law ($R_{\rm V}=3.1$), 
the relation of total gas column density and extinction is
$N(H + H_2)/A_{\rm V}=2\times 10^{21}\,{\rm cm}^{-2}\,{\rm mag}^{-1}$. 
The range of extinction in the Taurus map with 10 stars per cell corresponds 
to a range in column density from $N(H + H_2)=9.2\times 10^{20}\,{\rm cm}^{-2}$  
to $N(H + H_2) = 3.9 \times 10^{22} \,{\rm cm}^{-2}$.

The power spectra of the spatial distribution of visual extinction have been 
computed based on the 2MASS extinction maps. If the visual extinction is 
proportional to the total column density, these are also the power spectra 
of projected density. The power spectra are shown in Figure~\ref{f2}, for the 
maps with 3, 10 and 30 stars per cell. The noise power spectrum has been
removed, assuming uniform noise over the map. The statistical noise resulting
from the adaptive cell method is indeed uniform, because the number of stars
per cell is constant. With a uniform grid, instead, the noise would increase 
with increasing extinction, as the number of stars per cell decreases, so the
removal of the noise spectrum would be less trivial. Figure~\ref{f2} shows that 
i) the power spectrum is a power law, within a range of scales, 
ii) the slope of the power law slightly increases with decreasing spatial 
resolution (increasing number of stars per cell) and iii) at a comparable 
spatial resolution, the slope is significantly lower than the value of 
approximately -1.8 found by Padoan et al. (2004), based on J=1-0 $^{13}$CO 
observations and detailed radiative transfer modeling. 

The power spectrum derived from the $^{13}$CO maps in Padoan et al. (2004) was 
corrected for the effects of line saturation and spatial variations of kinetic 
temperature, based on three dimensional radiative transfer calculations of model 
clouds with realistic density distributions (obtained from simulations of supersonic 
turbulence). Only a small fraction of the mapped area corresponds to 
$A_{\rm V}>10$~mag, at the resolution of the $^{13}$CO map, so molecular depletion  
is not expected to affect significantly the power spectrum, as shown in Section~4.3.
Furthermore, Padoan et al. (2004)
studied two more regions, Perseus and Rosette, and found almost the same slope 
of approximately -1.8, which is also the power law predicted by simulations of 
supersonic turbulence with the same rms Mach number as found in those cloud 
complexes. Therefore the 2MASS extinction maps may underestimate the slope of the 
projected density power spectrum. This difference is presumably due to the presence 
of small scale intensity fluctuations of larger amplitude in the extinction map 
than in the $^{13}$CO map.

\section{Numerical Simulations and Synthetic Extinction Maps}

To verify that the extinction map method can reproduce the
correct power spectrum slope, we have generated and analyzed synthetic data.
We first generate a realistic three dimensional cloud model and a realistic 
sample of background stars with visual extinction determined by the column 
density of the cloud model at the positions of the stars. Synthetic extinction
maps are then computed in the same way as with the 2MASS stars.

The model density field is based on a simulation of supersonic turbulence,
with the ENZO code, using a direct Eulerian formulation of the Piecewise Parabolic 
Method (Colella \& Woodward 1984) to solve the equations of gas dynamics. 
In order to mimic the conditions found in molecular clouds, a quasi-isothermal 
equation of state was used, with the ratio of specific heats = 1.001. 
To maintain the turbulent kinetic energy in the computational box at a given 
level, we use a large-scale solenoidal force per unit mass with a fixed spatial 
pattern and a constant power in the range of wavenumbers $1<k<2$. 

Because real molecular clouds are known to be fragmented on very small scales and
each extinction measurement samples an extremely small cloud area
(the apparent projected area of the star), it is important to use a model density
field with a very large range of scales. For this reason, we have used our recent
turbulent simulation run on a uniform mesh of $1,024^3$ computational zones 
(Kritsuk et al. 2006, in preparation), which is the largest numerical experiment of 
highly supersonic turbulence carried out to date. The rms Mach number of the flow in this 
experiment was approximately 6. The density field was rescaled to correspond to an rms 
Mach number of 10 (comparable to the estimated Mach number in the Taurus region) 
according to the prescription in Padoan et al. (1998), which maintains the Lognormal 
form of the density histogram. A supersonic simulation with an rms Mach number of 10 and 
including the magnetic field would be more appropriate for this study. However,
for the purpose of generating synthetic extinction measurements it is more 
important to include a very large range of turbulent scales than the precise physical
parameters of the turbulent flow, and so we have chosen to use our largest available
run, despite the lack of the magnetic field and the Mach number value less than 10.

\begin{figure}[ht]
\centerline{
\epsfxsize=8.6cm \epsfbox{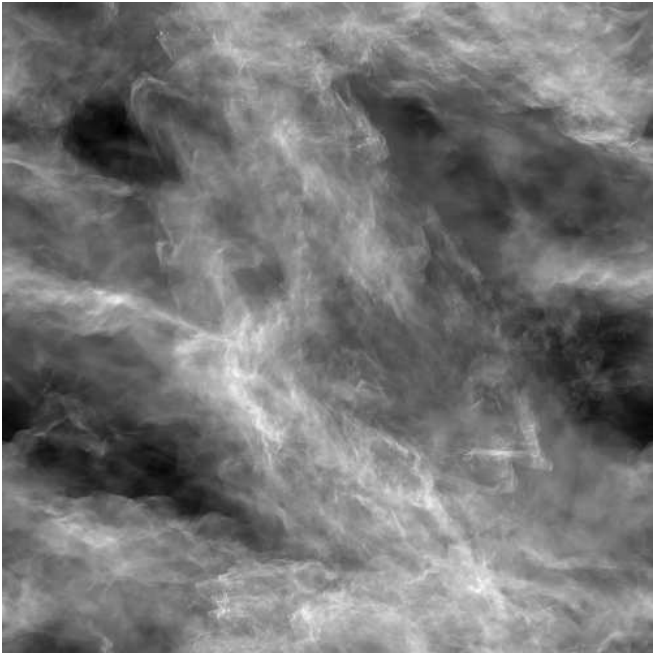}
}
\caption[]{Projected density from the $1,024^3$ simulation of supersonic
turbulence used to simulate the extinction measurements.
}
\label{f3}
\end{figure}

\begin{figure}[ht]
\centering
\epsfxsize=7.22cm \epsfbox{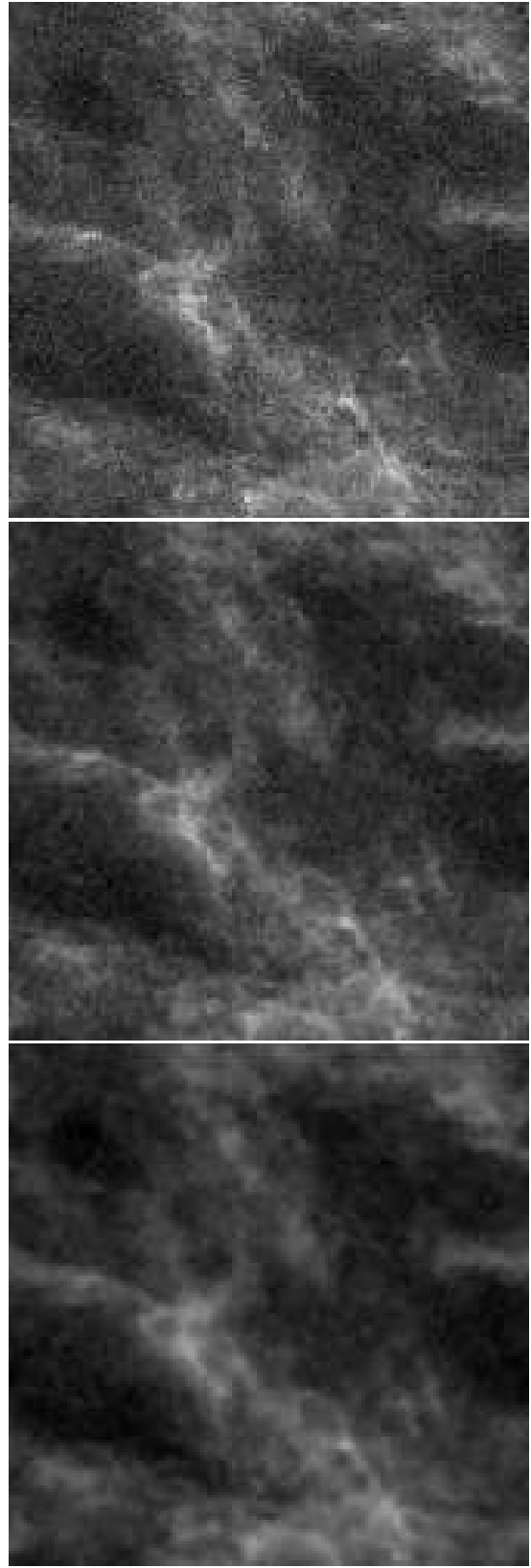}
\caption[]{Synthetic extinction maps of the projected density field shown
in Figure~\ref{f3} obtained with a simulated population of $10^5$ 
background stars (approximately the same number of stars used for the Taurus 
map). The maps are computed with 3, 10 and 30 stars per cell, from top to bottom.}
\label{f4}
\end{figure}

Three stellar catalogs were then generated with $10^4$, $10^5$ and $10^6$
stars. A $K$ luminosity function slope of 0.33 ($lg N=0.33 K + {\rm const}$)
was imposed, while the $H$ band magnitudes were randomly generated assuming
$\langle(H-K)\rangle=0.15$~mag and $\sigma(H-K)=0.12$~mag. The stars were uniformly
distributed in space with reddening proportional to the column density of the model 
cloud at the position of each star. To simulate the variation of stellar density with 
extinction, a magnitude limit was applied, which mimics the magnitude limit of the 2MASS
survey. Extinction maps were finally computed on these synthetic datasets with the same 
method used to compute the extinction maps from the 2MASS data.
The projected density of the cloud model is shown in Figure~\ref{f3}. Figure~\ref{f4}
shows the synthetic extinction maps from the catalogs with $10^5$ stars (comparable
to the number of stars used for the Taurus region here analyzed). The maps are 
based on 3, 10 and 30 stars per cell, from top to bottom. 

\begin{figure}[ht]
\centering
\epsfxsize=8.7cm \epsfbox{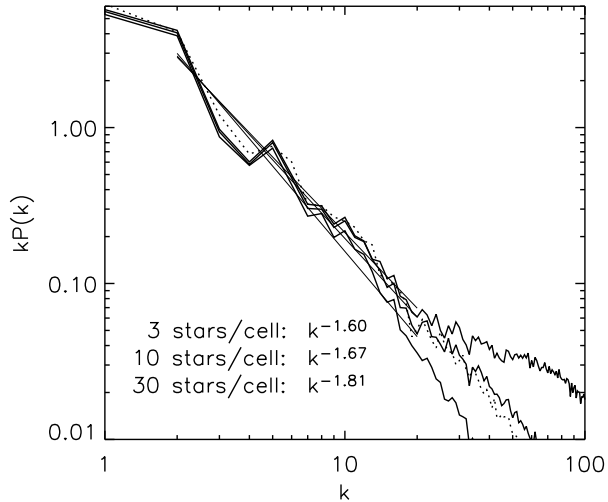}
\caption[]{Power spectra of the projected gas density field from Figure~\ref{f3} 
(dotted line) and of the three synthetic extinction maps shown in Figure~\ref{f4}.}
\label{f5}
\end{figure}

The power spectra of these synthetic extinction maps (assuming the dust accurately 
traces the gas) and of the projection of the original density field, are shown in 
Figure~\ref{f5}. As in the extinction
maps of the Taurus region, the power law slope of the power spectrum increases 
with decreasing spatial resolution (this trend is stronger for maps based on a 
lower number of stars). The slope remains close to the correct value 
for the original cloud model, particularly for the map with 10 stars per cell.
The variation of the slope with resolution is due to a gradual
smoothing of density structures. The column density is very filamentary, both
in the observations and in the simulations. When the density contrast of long coherent 
filaments is decreased in amplitude by the spatial smoothing, the power spectrum is 
slightly affected on all scales. Figure~\ref{f5} shows that the synthetic data 
closely reproduces the correct power spectrum, particularly for the map with 10 
stars per cell. Because the synthetic map is computed with approximately the 
same total number of stars as in the Taurus map, we assume that the observational 
map with 10 stars per cell should also closely match the correct power spectrum.
Furthermore, the angular resolution of the $^{13}$CO map of Taurus is intermediate
between that of the extinction maps with 10 and 3 stars per cell, so it is the power
spectra from these extinction maps that should be compared with the power spectrum 
of the CO map. The power spectrum slope for the dust is therefore approximately -1.2, 
much lower than the value of approximately -1.8 inferred from the CO emission.

\section{Discussion}

Radiative transfer effects and spatial variations of the gas temperature 
due to the complex cloud structure were accounted for by Padoan et al. (2004)
in deriving the power spectrum slope of the $^{13}$CO map of the Taurus, Rosette and
Perseus molecular cloud complexes. Depletion was estimated to be of secondary
importance at the range of $A_{\rm V}$ of the Taurus map. In their study of the
Chamaeleon complex, Hayakawa et al. (2001) found evidence of variations of $^{13}$CO 
versus $A_{\rm V}$ from cloud to cloud, which they attributed to chemical
fractionation increasing the $^{13}$CO abundance in regions with $2<A_{\rm V}<4$~mag.
Although their result was not confirmed by Kainulainen et al. (2006), CO abundance
variations may exist and may complicate the statistical analysis of CO maps. 
As Padoan et al. (2004) assumed a constant $^{13}$CO abundance in their models,
it is possible that the $^{13}$CO power spectra they derive are affected by 
variations in the $^{13}$CO abundance. However, the three-dimensional spatial
distribution of the $^{13}$CO abundance in a young turbulent cloud complex such 
as Taurus cannot be estimated with any degree of confidence, unless the chemical
model is coupled with the fluid equations and computed within a large simulation
of supersonic turbulence. In the absence of such a computation, we prefer to
assume a constant $^{13}$CO abundance in the approximate range $2<A_{\rm V}<10$~mag
of the $^{13}$CO map of the Taurus region. 

In the following we discuss two possible sources of the difference between the
extinction and the $^{13}$CO maps. First we consider the variation of the extinction 
efficiency due to grain growth toward regions of larger density. If grain
growth occurs and is not accounted for, variations in the near-infrared
extinction curve could cause an overestimate of the dust column density in 
regions of large extinctions. Second, we study the effect of variations of 
CO abundance, due to the formation and depletion of CO molecules, on the gas 
power spectrum. We finnd that variations in extinction efficiency or CO abundance 
are unlikely to cause very significant variations of the power spectra. 

We then briefly discuss preliminary results of numerical simulations of the 
clustering of heavy particles in turbulent  flows, which suggest the observed 
power spectrum discrepancies may originate from intrinsic differences between 
the spatial distributions of dust and gas, particularly on small scales. 
Finally, we address the role of the magnetic  field strength.

\subsection{Grain Growth and NIR Extinction Efficiency}

As explained above, in the extinction map method the NIR color 
excess, $E(H-K)$, is transformed into visual extinction, $A_V$, 
based on the Rieke and Lebofsky (1985) extinction law, giving the 
relation $A_{\rm V} = 15.87 \times E_{H-K_{\rm s}}$. It is believed 
that the assumption of a standard extinction law does not introduce 
significant errors in the column density determination, because the 
NIR reddening law has been found to have little variations in 
different lines of sight through the Galaxy, according to results reviewed 
in Mathis (1990). However, the old studies inferring a universal NIR 
extinction laws were all based on rather low-extinction data, for example
$A_V<2$~mag in Savage and Mathis (1979) and $A_V<5$~mag in Cardelli,
Clayton and Mathis (1989). A more recent study of the NIR extinction
curve at higher values of extinction by Moore et al. (2005) has 
found some evidence that the extinction curve tends to flatten with
increasing extinction, as predicted in the case of grain growth. 

These new results at higher extinction are not surprising, as it
is well established that grain reprocessing occurs at relatively large
density in interstellar clouds. Variations in the extinction parameter, 
$R_{\rm V}=A_{\rm V}/E(B-V)$, from the interstellar 
average of 3.1 to 4 or 5 in dark clouds and star-forming regions, are 
due to the flattening of the extinction curve at optical wavelengths. 
The flattening is interpreted as modifications of the population 
of small grains, due to ice mantle deposition and grain coagulation 
(e.g. Cardelli and Clayton 1991; Kim, Martin and Hendry 1993). 
Evidence of a decreased population of small grains due to grain coagulation, 
and formation of large fluffy aggregates has also been found at 
FIR wavelengths in dense filaments in Taurus (Stepnik et al. 2003), in the 
Polaris cirrus cloud (Bernard et al. 1999; Cambr\'{e}sy et al. 2001) and in a 
condensation of the Orion complex (Ristorcelli et al. 1998), thanks to a 
combination of IRAS and PRONAOS data. Grain growth with increasing density 
in dark clouds has also been probed by the increase in the wavelength of 
maximum polarization by absorption, interpreted as the result of ice mantle 
deposition in Taurus (Whittet et al. 2001) and grain coagulation in $\rho$ 
Ophiuchi (Vrba, Coyne and Tapia 1993). Finally, IR scattering also provides 
additional evidence of grain processing (e.g. Castelaz et al. 1985; 
Yamashita et al. 1989).

\begin{figure}[ht]
\centering
\epsfxsize=8.6cm \epsfbox{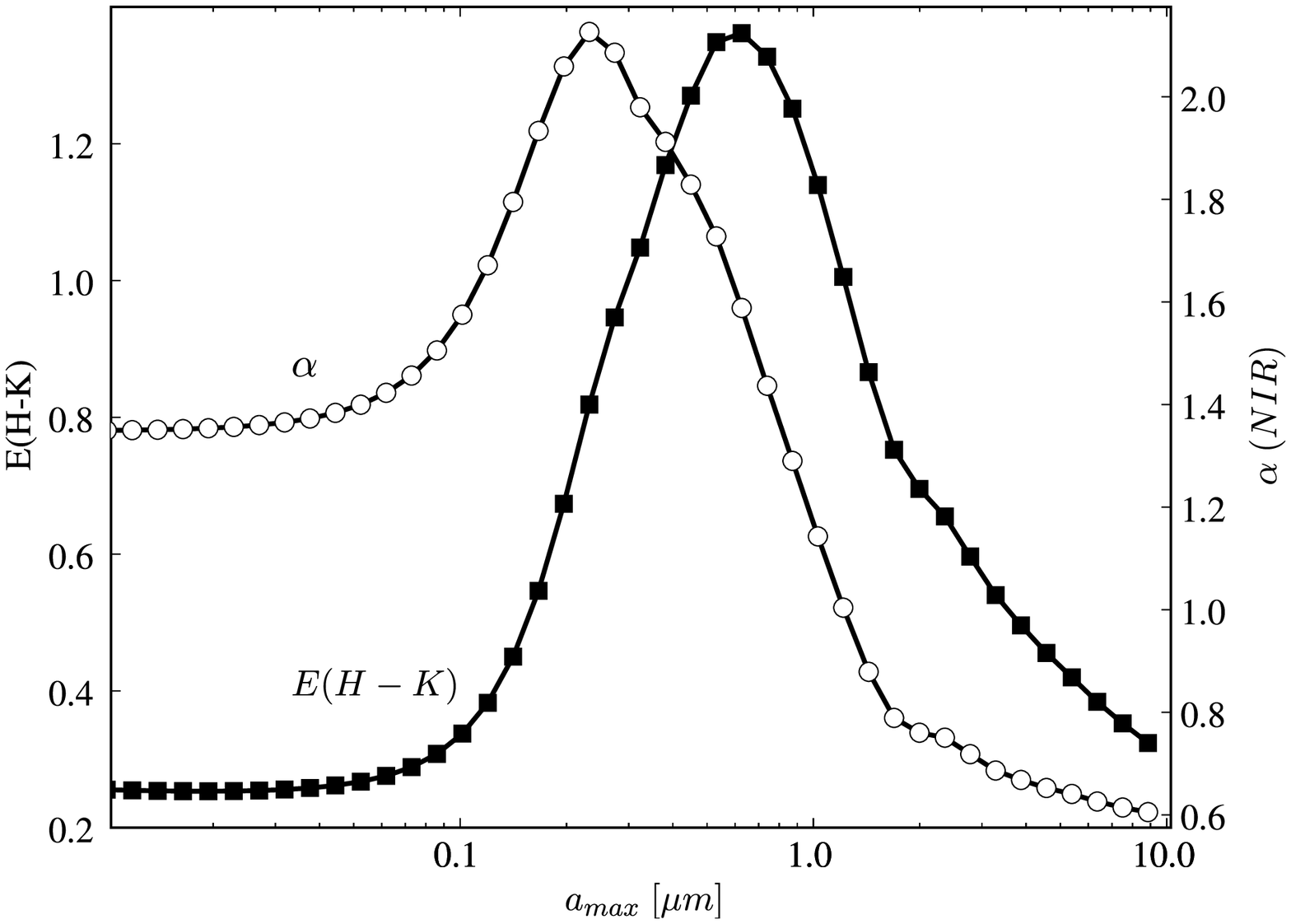}
\caption[]{NIR color excess $E(H-K)$ and spectral index $\alpha$ 
as a function of the maximum grain size, $a_{\rm max}$, for a MRN type 
grain size distribution, $n(a)\propto a^{-3.5}$ and a mixture of 
silicate and graphite grains with ratio of grain numbers Si:Gr equal
to 1.8. The minimum grain size is constant, $a_{\rm min}$=0.5\,nm. 
The total grain number is varied with $a_{\rm max}$ to maintain the 
total grain volume constant. The plotted values of $E(H-K)$ correspond 
therefore to a fixed value of dust column density and are scaled so 
that $E(H-K)=1$ for $a_{\rm max}=0.3 \mu$m. The spectral index is 
calculated between the $H$ and $K$ bands, 
$\alpha = -log(\tau_K/\tau_H)/log(2.22\mu m/1.65\mu m)$, 
using monochromatic values. The results are based on optical dust 
properties calculated by B. Draine and available at 
http://www.astro.princeton.edu/~draine/.}
\label{f6}
\end{figure}

The standard NIR extinction law corresponds to a power law 
$\tau \propto \lambda^{-\alpha}$, with $\alpha \approx 1.7-1.8$
(Cardelli et al. 1989; Martin and Whittet 1990). 
Moore et al. (2005) found a range of values of $\alpha$, between 1.11 
and 1.93, toward 9 ultracompact HII regions, with a clear trend of 
$\alpha$ to decrease with increasing extinction. This result indicates
that grain processing operates at large extinction and the NIR
extinction curves is therefore not universal as usually assumed based
on low extinction data. It is therefore possible that the extinction 
efficiency of dust grains varies in regions of large extinction,
affecting the conversion of NIR reddening into column density.
Indeed, the extinction efficiency is expected to increase, causing
an overestimate of the column density derived from NIR reddening,
if this effect is not accounted for.

In interstellar clouds, the typical size of dust grains affecting the NIR
extinction is approximately 0.2~$\mu$m. A modest grain growth by a factor 
of two would bring the 
typical size to 0.4~$\mu$m. This is precisely the range of sizes where 
the Rayleigh regime starts to be violated at the NIR bands, causing 
a significant increase of the scattering and extinction efficiency
(Kruegel and Siebenmorgen 1994). Figure~2 of Kruegel and Siebenmorgen 
(1994) shows that graphite grains can increase their extinction cross
section at a wavelength of 2.2~$\mu$m by almost an order of magnitude,
as the grain size increases from 0.2 to 0.4~$\mu$m. The variation in the 
extinction cross section of carbonaceous grains is instead much more 
modest. In the models of Kruegel and Siebenmorgen (1994) the maximum 
extinction is obtained with a grain radius of 1~$\mu$m, for silicate grains,
and a radius of 0.3~$\mu$m for carbonaceous grains.  

A comprehensive study of the modifications of grain chemical composition, 
density, shape and size with gas density in interstellar clouds and 
the resulting variations of the grains extinction efficiency is required
to accurately convert the observed NIR reddening into column density. 
Realistic grain models should be non-spherical (to explain the
observed polarization) and should contain a mixture of graphite, 
silicates, ices and empty space within each grain (Mathis and Whiffen 1989; 
Voshchinnikov and Mathis 1999). Such a study is beyond the scope of the present 
work. Here we only provide an example based on standard grain models. It is well 
known that the slope of the grain size distribution is constrained primarily by 
the value of $R_{\rm V}$, while the maximum grain radius, $a_{\rm max}$, is 
constrained by the UV extinction curve (Mathis 1990). Therefore, in Figure~\ref{f6},
we show, as an example, the variation of the NIR color excess, $E(H-K)$, 
as a function of $a_{\rm max}$ (for a fixed column density of dust), based on the 
standard MRN grain size distribution, $n(a)\propto a^{-3.5}$ (Mathis, Rumpl 
and Nordsieck 1977). 

In Figure~\ref{f6}, the minimum grain size 
is constant, $a_{\rm min}$=0.5\,nm, and we have adopted the optical dust 
properties calculated by Draine (http://www.astro.princeton.edu/$\sim$draine/).  
We plot the result for silicate and graphite grains alone and for a mixture 
of the two with the ratio of grain number Si:Gr=1.8. The corresponding values 
of the spectral index $\alpha$, calculated between the $H$ and $K$ bands, 
$\alpha = -log(\tau_K/\tau_H)/log(2.22\mu {\rm m}/1.65\mu {\rm m})$, using monochromatic 
values, are also plotted in the same figure. These values of $\alpha$ are 
consistent with those found by Moore et al. (2005), if the value of $a_{\rm max}$ 
for the mixture of graphite and silicate grains in the observed regions varies 
within the approximate range 0.3-1~$\mu$m as the density increases.\footnote{Notice
that Moore et al.'s study is for regions with $A_{\rm V}>15$~mag, so 
realistic variations of $\alpha$ in our range of values of $A_{\rm V}$ may be 
smaller. On the other hand, Moore et al.'s sightlines are several  
kpc. Although the total Av is high, a large fraction of the material may be  
unshielded and at low density.} 

Figure~\ref{f6} shows that $E(H-K)$ can grow at most by a factor of 4, for the 
mixture of silicate and graphite grains. This maximum variation of a factor of 4
is obtained if $a_{\rm max}$ grows from 0.1~$\mu$m in low density regions to 
0.6~$\mu$m in high density regions. We have estimated that such a variation 
could explain the shallow power spectrum of the extinction maps, as due to an 
increasingly overestimated cloud column density toward increasing density 
(hence $A_{\rm V}$ values). However, the required range of values for 
$a_{\rm max}$ is probably unrealistic, given the observational constraints 
from the extinction curve. The interstellar extinction curve of regions of 
low extinctions is well reproduced by an MRN grain size distribution of 
spheroidal grains with $a_{\rm max}=0.25$~$\mu$m (Gupta et al. 2005). 
If this is the value of $a_{\rm max}$ in regions of relatively low density, 
Figure~\ref{f6} shows that $E(H-K)$ can only increase with density by less 
than a factor of two, as the maximum grain size grows from $a_{\rm max}=0.25$~$\mu$m 
to $a_{\rm max}=0.6$~$\mu$m. In other words, neglecting the process of grain
growth can only cause errors of less than a factor of two in the dust column
density based on $E(H-K)$, which is insufficient to explain the difference 
between the gas and the dust power spectra.

\subsection{CO Formation and Depletion}

\begin{figure}[ht]
\centerline{
\epsfxsize=9.3cm \epsfbox{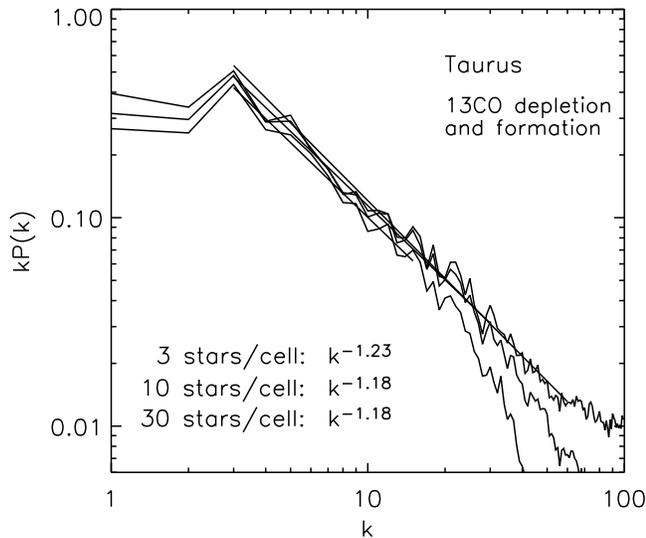}
}
\caption[]{Power spectra of the extinction maps from Figure~\ref{f1}, modified
by putting all values of $A_{\rm V} \le 3$~mag equal to zero (to test the effect
of a $^{13}$CO formation threshold) and all values of $A_{\rm V} \ge 10$~mag 
equal to 10~mag (to test the effect of depletion). The maps are re-sampled to 
the resolution of the map with 3 stars per cell as in Figure~\ref{f2}
}
\label{f8}
\end{figure}

Padoan et al. (2004) argued that CO depletion could not affect significantly their
estimate of the power spectrum of the $^{13}$CO map of the Taurus region, because
only a small fraction of the mapped area would correspond to $A_{\rm V}>10$~mag,
at the resolution of the Nagoya University survey. However, they could not quantify 
that statement, because there is no way to know a priori if larger column density
values are not observed because they are absent or because of depletion. Similarly, Padoan 
et al. (2004) argued that CO formation could not affect much the power spectrum,
because that would be relevant only for $A_{\rm V}<1$~mag. In reality, one
dimensional PDR models show that the $^{13}$CO line intensity becomes linear
with column density only around $A_{\rm V}>3$~mag (van Dishoeck \& Black 1988; 
Hollenbach \& Tielens 1997). From the $^{13}$CO map 
alone, there is no way to estimate this CO formation effect on the power spectrum, 
because we don't know if the absence of emission corresponds to very low total column 
density or to a low CO abundance.

However, the dust extinction map allows us to directly test the effect of CO
formation and depletion. If we assume the dust faithfully traces the total column
density, then the extinction map reveals the regions with $A_{\rm V}>10$~mag, where
depletion is expected to be important, and the regions with $A_{\rm V}<3$~mag, where
CO formation should be accounted for. A simple test of depletion is to recompute the
power spectrum of the extinction map after imposing that $A_{\rm V}=10$~mag in all
regions where we actually find $A_{\rm V}\ge 10$~mag. This mimics the fact that a 
$^{13}$CO map of the same region suffering from strong depletion would saturate
at approximately that value of $A_{\rm V}$. One would expect this effect to indeed
yield a steeper power spectrum, by smoothing out sharp small scale structures. 
A simple test for CO formation is to recompute the power spectrum of the extinction 
map after imposing that $A_{\rm V}=0$~mag in all regions where we actually find 
$A_{\rm V}<3$~mag. This may be an exaggerated way to mimic the effect of CO formation.
Nevertheless, we expect this effect to go in the direction of making the power spectrum 
shallower, by reducing the power in smooth large scale structures. If this is the case, 
accounting for CO formation can only increase the discrepancy between dust and gas power 
spectra, and a detailed modeling of that effect could not change the conclusion of this paper,
and is therefore not necessary for our present purposes.

The result of the ``depletion test'' is that the slopes of the power spectra 
corresponding to those in Figure~\ref{f2} change by an insignificant amount,
from -1.20, -1.26, and -1.41 to -1.25, -1.31, and -1.41 (the change is very small
even if the threshold is taken to be 7~mag instead of 10~mag). This confirms that, 
at a resolution of a few arc-minutes in the Taurus region, CO depletion can only 
have minor effects on the power spectrum of a $^{13}$CO map, as regions with 
$A_{\rm V}\ge 10$~mag are rare, small and barely resolved. Furthermore, it is well
understood that extinction maps tend to systematically underestimate the extinction 
in the regions of largest column density, as such regions can be detected only if 
a particularly bright stars is present in the background. We don't even try to 
correct for this bias, as it can only increase the already large discrepancy
between the gas and the dust.

The ``CO formation test'' goes in the predicted direction as well; it makes the
power spectrum shallower, which would further increase the discrepancy with
the gas power spectrum. The slopes change from -1.20, -1.26, and -1.41 to -1.20,
-1.15, and -1.18. In other words, if regions with $A_{\rm V}<3$~mag had to 
contain a $^{13}$CO abundance much below the dark cloud value, the CO power 
spectrum, based on the extinction map, would be expected to yield a slope
of approximately -1.20, instead of the estimated value of approximately -1.8.
CO formation, therefore, is not a candidate to explain the observed discrepancy 
either. 

The combined result of the two tests is illustrated in Figure~\ref{f8}. 
The power spectrum slope one would predict for the CO map, 
based on the dust extinction map (assumed to faithfully trace the total column 
density), are -1.23, -1.18, and -1.18, completely inconsistent with the 
$^{13}$CO observations.

\begin{figure}
\centering
\epsfxsize=9cm \epsfbox{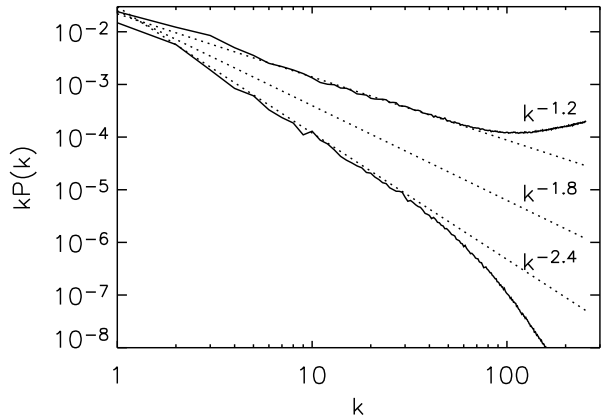}
\caption[]{Power spectrum of projected gas density (bottom solid line) and
projected particle (dust) density (top solid line) from a simulation of $512^3$
inertial particles embedded in a transonic (rms Mach number unity) hydrodynamic 
turbulent flow simulated on a mesh of $512^3$ computational zones. The particle
power spectrum is for particles with friction time approximately twice the
turbulence dynamical time at the Kolmogorov dissipation scale. Three power laws
(dotted lines) are also plotted for comparison. The $k^{-1.8}$ power law corresponds
roughly to the gas projected density power spectrum in supersonic and 
super-Alfv\'{e}nic simulations with rms Mach number 10. The particle power
spectrum is much shallower than the gas power spectrum on all scales up to the
outer scale of the turbulence (the box size or $k=1$), as the result of the
spatial clustering of the particles.}
\label{f7}
\end{figure}

\subsection{Clustering of Dust Grains by the Cloud Turbulence}

If the power spectrum of the dust column density is indeed shallower
than the power spectrum of the gas column density, fluctuations in the dust to 
gas ratio must be present, with amplitude increasing toward smaller scales. 
Such fluctuations are indeed the main effect of 
the clustering of dust grains due to the cloud turbulence. A detailed discussion
of this process in the context of molecular clouds is given in Padoan et al. 
(2006, in preparation). In summary, turbulent flows advect heavy particles by gas 
drag and concentrate them in regions of low flow vorticity, because the particles 
are centrifugally expelled from vorticity tubes (Maxey 1987). The amplitude of the 
clustering grows toward smaller scale, all the way down to the scale of Brownian 
diffusion. As a result, fluctuations of the dust to gas ratio are created, and 
can reach a few orders of magnitude at the Brownian scale, of order $10^{11}$~cm. 
The amplitude of the clustering depends also on the particle size, with the largest 
values for particles with friction timescale equal to the Kolmogorov timescale 
(the dynamical time of the turbulence at the Kolmogorov dissipation scale). 

The relative motion of such particles is controlled primarily by the gas 
turbulence at the Kolmogorov dissipation scale, of order $10^{14}$~cm. However, 
preliminary results in Padoan et al. (2006, in preparation) show that the 
particles cluster on all scales, with amplitude decreasing with increasing scale. 
As a result, the projected density of particles exhibit a shallower power 
spectrum than the projected density of the turbulent gas. The power spectra
of both gas and particle projected density are shown in Figure.~\ref{f7} (lower
and upper solid lines respectively). A difference between particle 
and gas power spectra on all scales is obtained for particles with friction 
time of the order of the Kolmogorov time (in Figure.~\ref{f7} the particle power 
spectrum is shown for particles with friction time equal to twice the Kolmogorov time).
For typical physical parameters of molecular cloud, this corresponds to dust grains 
of roughly the same size as the grains that dominate the NIR extinction resulting 
in the 2MASS maps.

The simulations of Padoan et al. (2006, in preparation) compute the trajectories 
of $512^3$ inertial particles embedded in a turbulence flow simulated on a mesh 
of $512^3$ computational cells. The turbulent flow is transonic and is not 
magnetized. Although the slope of the power spectrum shown in Figure~\ref{f7} 
is roughly the same as the slope of the 2MASS extinction map of the Taurus molecular 
cloud, the simulations cannot be compared directly with the observations, as they 
are meant to represent a smaller scale ($<0.1$~pc), where the turbulence is not 
supersonic. The effect of magnetic fields can also be important in the particle 
transport, because the gyration time of the grains is typically shorter than their 
gas friction time (Yan, Lazarian and Draine 2004). However, our numerical result 
shows that grain clustering by the cloud turbulence can potentially be a dominant 
mechanism to generate small scale fluctuations of the dust to gas ratio and hence a 
power spectrum of dust column density shallower than the power spectrum of the gas 
column density. A similar mechanism may still operate in magnetized and highly 
supersonic turbulent flows.

\subsection{Average Magnetic Field Strength}

Padoan et al. (2004) concluded that the estimated
power spectrum slope in the Taurus region (in the Rosette and Perseus
molecular cloud complexes as well) is consistent with supersonic 
and super-Alfv\'{e}nic turbulence, while flow models with much 
stronger magnetic fields are ruled out, as they predict a shallower
power spectrum than observed. We are here referring to the value of the 
magnetic field strength volume averaged over the entire Taurus molecular
cloud complex. Individual cores are likely to have an enhanced field
strength much larger than the averaged value, as a natural result of 
shock compressions in super-Alfv\'{e}nic turbulence. This idea was first
proposed by Padoan and Nordlund (1999), and is still considered 
controversial by some researchers, due to the difficulty of directly
mapping the magnetic field strength in molecular clouds.

The shallower power spectrum of the dust extinction map compared with the
CO map may be taken as evidence that the average magnetic field strength
in the Taurus region is stronger than inferred by Padoan et al. (2004).
However, this conclusion would be incorrect for the following reason.   
The combined analysis of the present work and of Padoan et al. (2004) 
does not leave much space to believe that the $^{13}$CO power spectrum 
we derive is greatly unreliable and does not trace at all the total 
column density. Once major radiative transfer, chemistry and depletion 
effects are ruled out on the scales of interest, we are truly compelled 
to question our understanding of interstellar dust grains, their composition 
and optical properties, their formation and evolution, and their dynamics in 
a turbulent magnetized gas. In the light of these many uncertainties still 
characterizing the physics of interstellar dust, it is legitimate to ask 
``Can we trust the dust?'' and to propose that the agreement between the 
power spectrum predicted by super-Alfv\'{e}nic turbulence and the observed 
one is not a mere coincidence.

Finally, the main source of the three dimensional temperature and radiative 
transfer effects discussed in Padoan et al. (2004), and of the results of
the synthetic dust extinction modeling presented here, is the strong 
inhomogeneity of the density field in supersonic turbulence. The detailed 
statistics of such a density field and their dependence on the magnetic 
field strength could only cause small modifications to these analysis. 
In this work we have used a purely hydrodynamical simulation because
it is the largest simulation of supersonic turbulence to date, and the 
presence of a huge range of scales of turbulent density fluctuations is
far more important for simulating the extinction measurements than the 
detailed statistics of that density field.

\section{Conclusions}

We have studied the power spectrum of NIR extinction maps of the Taurus region
and found that it is significantly shallower than the power spectrum obtained 
by Padoan et al. (2004) from a $^{13}$CO map of the same region.
The extinction map has a power spectrum slope of approximately -1.2,
while the gas map has a slope of approximately -1.8. The $^{13}$CO study in Padoan
et al. (2004) relied on detailed modeling of the observations based on three 
dimensional simulations of supersonic turbulence and radiative transfer 
calculations. Here we have further ruled out other uncertainties, related 
to the processes of CO formation and depletion, as a possible explanation
for the observed discrepancy. The present extinction study
relies on three dimensional simulations of supersonic turbulence as well,
in order to reproduce the observations and compute synthetic extinction maps.
The synthetic extinction maps confirm that the power spectrum derived
from the observations should correspond to the power spectrum of the actual spatial
distribution of the dust, if the extinction is proportional to the dust column density.

The discrepancy between the power spectrum slope of the dust and that of the gas 
could be in part understood if the dust column density were increasingly
overestimated towards larger extinctions, due to the effect of grain growth 
on the grain extinction efficiency. We cannot entirely rule out this 
explanation, given the still limited knowledge of dust grain properties 
(different grain models can satisfy the same observational constraints, 
and most grain models are not realistic enough, as they overestimate the 
abundance of carbon and silicon relative to ISM values in order to explain 
the extinction curve --Gupta et al. 2005). However, based on available models,
the UV extinction curve is best fit by a value of $a_{\rm max}=0.25$~$\mu$m, 
in regions of relatively low extinction. Starting from that size, grain growth
can only affect the NIR color excess by less than a factor of two, assuming that
grains really grow up to a value $a_{\rm max}=0.6$~$\mu$m in regions of high
extinction. In order to explain the power spectrum slope, the NIR excess should
vary by at least a factor of 4, which would be possible only with variations
in $a_{\rm max}$ from 0.1 to 0.6~$\mu$m. Unless evidence is found in favor of 
values of $a_{\rm max}<0.25$~$\mu$m in low extinction regions of Taurus, or 
unless the available grain models are found to be very unreliable (for example
due to the formation of composite fluffy aggregates), our results
suggest the existence of intrinsic spatial fluctuations of the dust to gas 
ratio, with amplitude increasing toward smaller scale.

The comparison between the dust and the gas maps of the Taurus region 
presented in this work may be further improved in the near future, based
on new observational data. Goldsmith et al. (2005) have recently carried out 
a new survey of the Taurus region, combining $^{13}$CO and $^{12}$CO spectra from 
the FCRAO 14~m telescope. The better resolution of this survey compared with 
the survey from the Nagoya University 4~m telescope (Mizuno et al. 1995) used in 
Padoan et al. (2004), and the increased range of column density from the 
combination of $^{12}$CO and $^{13}$CO spectra, will yield a much improved gas 
column density map of the Taurus region. One could also measure the spectral type 
of background stars, at least for positions of unusually large 2MASS extinction 
relative to CO column density, to reduce the uncertainty of the extinction 
measurements.

Because grain clustering is expected to be very strong
on small scales, high resolution studies of the dust spatial distribution
should be well suited to examine this process. Padoan et al. (2006) and
Juvela et al. (2006) have extensively investigated a new method of mapping
dark clouds by observing their NIR scattered light. Previous observations
have already successfully detected dark clouds through the NIR scattering 
of the normal ISRF (Lehtinen and Mattila 1996; Nakajima et al. 2003; 
Foster and Goodman 2006). The combination of relatively short wavelength, 
large ground based telescopes and large NIR cameras will make it possible to map
vast areas of interstellar clouds at an unprecedented arc-second resolution, 
with an estimated uncertainty in column density below 20\% in the range 
$1<A_{\rm V}<20$~mag, based on the simultaneous use of the J, H and K bands 
(Juvela et al. 2006). NIR scattering maps may therefore provide 
evidence of dust segregation by directly imaging the spatial clustering
of dust grains.

Finally, the process of inertial particle clustering has so far been 
confirmed with simulations of incompressible or transonic hydrodynamic
turbulence. A similar process may occur also in magnetized and highly 
supersonic turbulent flows such as those found in molecular clouds,
and future simulations of inertial particles in this regime of turbulence 
will be necessary for a direct comparison with observations of parsec scale 
regions.

\acknowledgements
The research of WDL was conducted at the Jet Propulsion Laboratory, 
California Institute of Technology under support from the National 
Aeronautics and Space Administration (NASA).  WDL and PP had partial 
support for this research under a grant from NASA's Astrophysics 
Data Program (ADP). M.J. acknowledges the support of the Academy of Finland
Grants no. 206049 and 107701. We would like to thank Dr. Karen Willacy for 
useful discussions during the course of this work, and for reading and 
commenting on the manuscript, and the referee for many useful comments
and corrections.

\bibliographystyle{apj}
%\bibliography{apj-jour,cosmic,MC,padoan}

%
%Taurus
%        nb   sig   res   AVmax
%        100  0.37  12.3   8.0
%         30  0.40   6.7  11.7
%         10  0.46   3.9  19.5
%          3  0.58   2.3  26.3
%          1  0.76   1.5  32.7
%Perseus
%        nb   sig   res   AVmax
%        100  0.30  12.0  10.1
%         30  0.33   6.5  13.5
%         10  0.41   3.7  19.6
%          3  0.57   1.9  27.0
%Serpens
%        nb   sig   res   AVmax
%        100  0.71   4.2  14.7
%         30  0.73   2.3  22.4
%         10  0.78   1.3  28.2
%          3  0.92   0.7  37.6
%          1  1.11   0.3  55.9

\end{document}